\documentclass[aps,prd,twocolumn,superscriptaddress,showpacs,preprintnumbers,amsmath,amssymb]{revtex4}

\usepackage{graphicx} 
\usepackage{dcolumn}  

\graphicspath{{ps}}

\begin{document}

\title{
\boldmath {
 \quad\\[1.0cm] Search for $B \to \phi \pi$ decays} }

\affiliation{University of Bonn, Bonn}
\affiliation{Budker Institute of Nuclear Physics SB RAS and Novosibirsk State University, Novosibirsk 630090}
\affiliation{Faculty of Mathematics and Physics, Charles University, Prague}
\affiliation{University of Cincinnati, Cincinnati, Ohio 45221}
\affiliation{Department of Physics, Fu Jen Catholic University, Taipei}
\affiliation{Gifu University, Gifu}
\affiliation{Hanyang University, Seoul}
\affiliation{University of Hawaii, Honolulu, Hawaii 96822}
\affiliation{High Energy Accelerator Research Organization (KEK), Tsukuba}
\affiliation{Indian Institute of Technology Guwahati, Guwahati}
\affiliation{Indian Institute of Technology Madras, Madras}
\affiliation{Institute of High Energy Physics, Chinese Academy of Sciences, Beijing}
\affiliation{Institute of High Energy Physics, Vienna}
\affiliation{Institute of High Energy Physics, Protvino}
\affiliation{Institute for Theoretical and Experimental Physics, Moscow}
\affiliation{J. Stefan Institute, Ljubljana}
\affiliation{Kanagawa University, Yokohama}
\affiliation{Institut f\"ur Experimentelle Kernphysik, Karlsruher Institut f\"ur Technologie, Karlsruhe}
\affiliation{Korea Institute of Science and Technology Information, Daejeon}
\affiliation{Korea University, Seoul}
\affiliation{Kyungpook National University, Taegu}
\affiliation{\'Ecole Polytechnique F\'ed\'erale de Lausanne (EPFL), Lausanne}
\affiliation{Faculty of Mathematics and Physics, University of Ljubljana, Ljubljana}
\affiliation{Luther College, Decorah, Iowa 52101}
\affiliation{University of Maribor, Maribor}
\affiliation{Max-Planck-Institut f\"ur Physik, M\"unchen}
\affiliation{University of Melbourne, School of Physics, Victoria 3010}
\affiliation{Graduate School of Science, Nagoya University, Nagoya}
\affiliation{Kobayashi-Maskawa Institute, Nagoya University, Nagoya}
\affiliation{Nara Women's University, Nara}
\affiliation{National Central University, Chung-li}
\affiliation{Department of Physics, National Taiwan University, Taipei}
\affiliation{H. Niewodniczanski Institute of Nuclear Physics, Krakow}
\affiliation{Nippon Dental University, Niigata}
\affiliation{Niigata University, Niigata}
\affiliation{University of Nova Gorica, Nova Gorica}
\affiliation{Osaka City University, Osaka}
\affiliation{Pacific Northwest National Laboratory, Richland, Washington 99352}
\affiliation{Panjab University, Chandigarh}
\affiliation{Research Center for Electron Photon Science, Tohoku University, Sendai}
\affiliation{University of Science and Technology of China, Hefei}
\affiliation{Seoul National University, Seoul}
\affiliation{Sungkyunkwan University, Suwon}
\affiliation{School of Physics, University of Sydney, NSW 2006}
\affiliation{Tata Institute of Fundamental Research, Mumbai}
\affiliation{Excellence Cluster Universe, Technische Universit\"at M\"unchen, Garching}
\affiliation{Tohoku Gakuin University, Tagajo}
\affiliation{Tohoku University, Sendai}
\affiliation{Department of Physics, University of Tokyo, Tokyo}
\affiliation{Tokyo Institute of Technology, Tokyo}
\affiliation{Tokyo Metropolitan University, Tokyo}
\affiliation{Tokyo University of Agriculture and Technology, Tokyo}
\affiliation{CNP, Virginia Polytechnic Institute and State University, Blacksburg, Virginia 24061}
\affiliation{Wayne State University, Detroit, Michigan 48202}
\affiliation{Yamagata University, Yamagata}
\affiliation{Yonsei University, Seoul}
 \author{J.~H.~Kim}\affiliation{Sungkyunkwan University, Suwon}\affiliation{Korea Institute of Science and Technology Information, Daejeon} 
\author{M.~Nakao}\affiliation{High Energy Accelerator Research Organization (KEK), Tsukuba} 
  \author{I.~Adachi}\affiliation{High Energy Accelerator Research Organization (KEK), Tsukuba} 
  \author{K.~Adamczyk}\affiliation{H. Niewodniczanski Institute of Nuclear Physics, Krakow} 
  \author{H.~Aihara}\affiliation{Department of Physics, University of Tokyo, Tokyo} 
  \author{D.~M.~Asner}\affiliation{Pacific Northwest National Laboratory, Richland, Washington 99352} 
  \author{V.~Aulchenko}\affiliation{Budker Institute of Nuclear Physics SB RAS and Novosibirsk State University, Novosibirsk 630090} 
  \author{T.~Aushev}\affiliation{Institute for Theoretical and Experimental Physics, Moscow} 
  \author{A.~M.~Bakich}\affiliation{School of Physics, University of Sydney, NSW 2006} 
  \author{K.~Belous}\affiliation{Institute of High Energy Physics, Protvino} 
% ...
\collaboration{The Belle Collaboration}


\begin{abstract}
{We report on a search for the charmless decays $B^{+} \to\phi\pi^{+}$ and $B^{0} \to\phi \pi^{0}$ that are strongly suppressed in the Standard Model. The analysis is based on a data sample of $657 \times 10^6$ $B \overline{B}$ pairs collected at the $\Upsilon(4S)$ resonance with the Belle detector at the KEKB asymmetric-energy $e^+ e^-$ collider. We find no significant signal and set upper limits of $3.3 \times 10^{-7}$ for $B^{+} \to \phi \pi^{+}$ and $1.5 \times 10^{-7}$ for $B^0 \to \phi \pi^0$ at the 90$\%$ confidence level.}

\end{abstract}


\maketitle

\tighten

{\renewcommand{\thefootnote}{\fnsymbol{footnote}}}
\setcounter{footnote}{0}

{In the Standard Model (SM), the charmless two-body hadronic decays $B^{+} \to\phi\pi^{+}$~\cite{conjugate} and $B^0 \to\phi \pi^0$ are highly suppressed since they are forbidden at tree level and are only possible through the penguin process shown in Fig.~{\ref{feynman}}(a). The expected SM branching fractions for these decays are ${\cal B}(B^{+} \to \phi \pi^+) \sim 3.2 \times 10^{-8}$ and ${\cal B}(B^0 \to \phi \pi^0 ) \sim 6.8 \times 10^{-9}$~\cite{SMth}, in which the largest contribution comes from radiative corrections and $\omega$-$\phi$ mixing. In some New Physics (NP) scenarios such as models with a $Z'$ boson ~\cite{th1,th22} or the Constrained Minimal Supersymmetric Standard Model (CMSSM)~\cite{th3}, the branching fractions could be enhanced up to the $10^{-7}$ level. Figure~{\ref{feynman}}(b) shows a typical CMSSM contribution to $B \to \phi \pi$.

\begin{figure}[htb!]
\centering{
\includegraphics[width=0.48\textwidth]{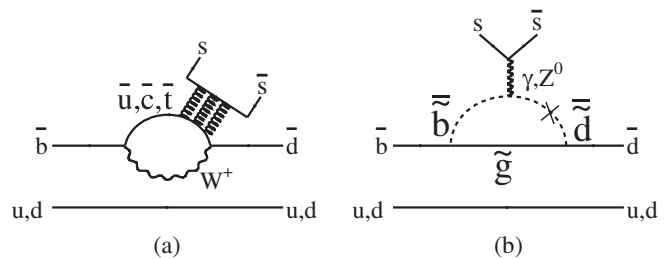}
}
\caption{(a) The SM three-gluon hairpin penguin diagram for $B \to \phi \pi$ decays. (b) One of the CMSSM diagrams that contributes to $B \to \phi \pi$. In both a) and b), the $s \bar{s}$ quark pair hadronizes as a $\phi$ meson.
{\bf  } }
\label{feynman}
\end{figure}
Since $B \to \phi \pi$ decays are very sensitive to NP, measurements of these decays may constrain and potentially reveal such contributions. Furthermore, measurements of $B \to \phi \pi$ decays also provide a means to study SM contributions from suppressed diagrams in other important decay modes such as $B^0 \to \phi K^0$~\cite{ex1}. A previous search by the ${\rm BaBar}$ collaboration set upper limits of $\mathcal{B}(B^{+} \to \phi \pi^{+}) < 2.4 \times 10^{-7}$ and $\mathcal{B}(B^0 \to\phi \pi^0) < 2.8 \times 10^{-7}$ at the 90\% confidence level (CL)~\cite{ex2}. A later measurement of the three-body inclusive branching fraction for $B^+ \to K^+ K^- \pi^+$~\cite{chargednonresonant} also did not report any evidence for $B^+ \to \phi \pi^+$.
 }\par

{In this paper, we report on a search for $B^{+} \to \phi \pi^{+}$ and $B^0 \to \phi \pi^0$ based on a $605 ~ \mathrm{fb^{-1}}$ data sample, which corresponds to ($657 \pm 9) \times 10^6$ $B \overline{B}$ events. The data were collected with the Belle detector~\cite{Belle} at the KEKB~\cite{KEKB} asymmetric-energy $e^+ e^-$ collider operating at the $\Upsilon(4S)$ resonance. }\par
{The Belle detector is a large-solid-angle magnetic spectrometer that consists of a silicon vertex detector (SVD), a 50-layer central drift chamber (CDC), an array of aerogel threshold Cherenkov counters (ACC),  
a barrel-like arrangement of time-of-flight
scintillation counters (TOF), and an electromagnetic calorimeter (ECL)
comprised of CsI(Tl) crystals located inside 
a superconducting solenoid coil that provides a 1.5~T
magnetic field.  An iron flux-return located outside of
the coil is instrumented to detect $K_L^0$ mesons and to identify
muons.  
Two inner detector configurations were used: a 2.0 cm beampipe
and a 3-layer silicon vertex detector were used for the first sample
of $152 \times 10^6 B\bar{B}$ pairs, while a 1.5 cm beampipe, a 4-layer
silicon detector, and a small-cell inner drift chamber were used to record  
the remaining $505 \times 10^6 B\bar{B}$ pairs~\cite{svd2}.  }\par
{To search for $B^{+} \to \phi \pi^{+}$ and $B^{0} \to \phi \pi^{0}$, we combine $\phi \to K^+ K^-$ candidates with either a $\pi^{+}$ or $ \pi^{0} \to \gamma \gamma$. Particle identification (PID) for charged kaons from the $\phi$ decays and the charged pion is based on the likelihood ratios $R_{K,\pi } = \frac { L_K}{L_K + L_{\pi}}$, where $L_K$ and $L_{\pi}$ denote, respectively, the individual likelihoods for kaons and pions derived from ACC and TOF information and $dE/dx$ measurements in the CDC. The PID selections, $R_{K,\pi } > 0.3$ for kaon candidates and $R_{K,\pi } < 0.2 $ for pion candidates, are applied to all charged particles. The PID efficiencies are 87\% (86\%) for kaon pairs (high momentum single pions) in $B^{+} \to \phi \pi^{+}$ and 86\% for kaon pairs in $B^{0} \to \phi \pi^{0}$, while the probability of misidentifying a kaon as a pion (a pion as a kaon) is 6\% (12\%) for both modes. Candidate $\pi^0$'s are reconstructed from $\gamma$ pairs that have invariant mass between 115.3 $\mathrm{MeV}/c^2$ and 152.8 $\mathrm{MeV}/c^2$, corresponding to $\pm 2.5 \sigma$ standard deviations ($\sigma$). In addition, these photons are required to have energies greater than 0.2 $\rm{GeV}$. A $K^{+} K^{-}$ pair is required to have an invariant mass within the range $1.008~{\rm{GeV}}/c^2 < M_{K^{+}K^{-}} < 1.031~{\rm{GeV}}/c^2$ ($\pm2.5$ times the $\phi$ full width).}\par
{$B$ meson candidates are identified with two kinematic variables: beam-energy-constrained mass, $M_{\rm {bc}} = \sqrt{E^2_{\rm{beam}} - |\sum_i \vec{p_i}|^2}$, and energy difference $\Delta E = \sum_i E_{i}  - E_{\mathrm{beam}}$, where $E_{\mathrm{beam}}$ is the beam energy, and $\vec{p_i} $ and $E_i$ are the momenta and energies, respectively, of the daughters of the reconstructed $B$ meson candidate in the center-of-mass (CM) frame. We fit $B$ candidates that lie within the fit region defined by $|\Delta E| <$ 0.1 $\mathrm{GeV}$ and $M_{\rm {bc}} >$ 5.20 $\mathrm{GeV}/c^2$ for $B^{+} \to \phi \pi^{+}$ and $ |\Delta E| <$ 0.4 $\mathrm{GeV}$ and $M_{\rm {bc}}>$ 5.20 $\mathrm{GeV}/c^2$ for $B^0 \to \phi \pi^0$. The signal regions are defined by $|\Delta E| <$ 0.04 $\mathrm{GeV}$ $(\pm 3.0 \sigma)$ and $M_{\rm {bc}} >$ 5.27 $\mathrm{GeV}/c^2$ $(\pm 3.0 \sigma)$ for $B^{+} \to \phi \pi^{+}$, and $-0.16$ $\mathrm{GeV}$ $ (5.0 \sigma)$ $< \Delta E <$ 0.10 $\mathrm{GeV}$ $ (3.0 \sigma)$ and $M_{\rm {bc}} >$ 5.27 $\mathrm{GeV}/c^2$ $(\pm 3.0 \sigma)$ for $B^0 \to \phi \pi^0$. We select an asymmetric signal region for $B^0 \to \phi \pi^0$ since photons may interact with the intervening detector material before entering the ECL and there may be energy leakage from the ECL crystals.}\par
{The main background arises from the continuum process, $e^{+} e^{-} \to q \bar{q}$, where $q = u, d, s, c$. To suppress this, observables based on the event topology are utilized. The event shape in the CM frame is spherical for $B\bar{B}$ events and jet-like for continuum events. This difference is exploited by the event-shape variable, which is a Fisher discriminant formed out of 16 modified Super Fox-Wolfram moments~\cite{fox,KSFW} calculated in the CM frame. The angle of the $B$ flight direction (${{\theta}^* _{B}}$) with respect to the beam axis provides additional discrimination since it is distributed as $(1-\cos^2\theta^{*}_B)$ for $B$ decays but flat for continuum. The distance in the $z$ direction ($\Delta z$) between the signal $B$ vertex~\cite{vertexres} and that of the other $B$ is used in the continuum suppression if $| \Delta z |$ is less than 2.0 $\rm{mm}$. For $B$ events, the average value of $| \Delta z |$ is approximately 0.2 $\rm{mm}$, whereas continuum events tend to have a common vertex that is measured with a resolution of about 1.0 $\rm{mm}$. In addition, the helicity angle ($\theta_H$) discriminates between the signal and continuum events, where $\theta_H$ is the angle between the final state $K^+$ direction and the $B$ meson direction in the $\phi$ rest frame. We first calculate the individual probability density function (PDF) for the Fisher discriminant, ${\rm {cos}}\theta^* _{B}$, $\Delta z$ and ${\rm{cos}}\theta_{H}$, and then obtain their product, 
\begin{equation}
L_{S(q\bar{q})} = {\prod}_i {L^{i}_{S(q\bar{q})}}, 
\end{equation}
where ${L^{i}_{S(q\bar{q})}}$ denotes the signal ($q\bar{q}$) likelihood of the continuum suppression variable $i$. The PDFs for signal, generic $B$, and continuum events are obtained from the GEANT3-based~\cite{MC} Monte Carlo (MC) simulation. The variable used for continuum suppression is the likelihood ratio ($R_S$) defined as  
\begin{equation}
R_S = \frac{L_S}{L_S + L_{q\bar{q}}}.
\end{equation}
Additional background suppression is achieved through the use of a $B$-flavor tagging algorithm~\cite{TaggingNIM}, which provides two outputs: $q = \pm 1$ indicating the flavor of the other $B$ in the event, and $r$, which takes a value between 0 and 1 and is the quality of the flavor determination. Events with a high value of $r$ are considered to be well-tagged. The continuum background is reduced by applying a $qr$-dependent selection requirement on $R_S$. This requrement is optimized in three $qr$ regions for $B^{+} \to \phi \pi^{+}$: $-1 \leq qr < -0.5$, $-0.5 \leq qr < -0.1$, and $-0.1 \leq qr \leq 1$. For $B^{0} \to \phi \pi^{0}$, since we do not distinguish the $B$ flavor, we use three $r$ intervals: $0 \le r < 0.25$, $0.25 \le r < 0.70$, and $0.70 \le r \leq 1$. The requirements are chosen to maximize a figure of merit (FOM) defined as 
\begin{equation}
FOM = \frac{N_{S}}{\sqrt{N_{S}+N_{B}}},
\end{equation}
where $N_S$ is the number of signal MC events in the signal region and $N_B$ is the number of background events estimated in the signal region by assuming $\mathcal{B}(B^{+} \to \phi \pi^{+} ) = 2.4 \times 10^{-7}$ and $\mathcal{B}(B^0 \to \phi \pi^0 ) = 2.8 \times 10^{-7}$. Our background suppression eliminates 99.4\% (99.7\%) of continuum background while retaining 55.8\% (43.9\%) of the signal events for $B^{+}\to\phi\pi^{+}$ ($B^{0}\to\phi\pi^{0}$).}\par
{Backgrounds from $B$ decays are studied using large MC samples. The sample size for charmless decays from $b \to u, d, s$ transitions corresponds to 50 times the data luminosity. For $B \to \phi \pi^+$, the $b \to s$ process $B \to \phi K^+$ is the dominant background, arising from kaon-to-pion misidentification. For $B^0 \to \phi \pi^0$, a decay with a $\pi^0$ in the final state such as $B^0 \to \phi K^0 _S$, is the dominant contribution. This background has a signal-like distribution in $M_{\rm {bc}}$. However, the background populates the negative $\Delta E$ region with small overlap with the signal, so its contribution can be extracted from a fit.}\par
{Signal yields for $B \to \phi \pi$ decays are obtained by performing a two-dimensional extended unbinned maximum likelihood (ML) fit to the observables $M_{\rm {bc}}$ and $\Delta E $. The likelihood is
\begin{equation}
L = e^{-{\sum_{i}^{} N_i }} \times \prod_{j}^{} \left[ \sum_{i} N_i P_i (M_{\rm bc}, \Delta E)_j \right],
\end{equation}
where the index $i$ denotes signal, continuum, $b \to c$ background, and $b\to u, d, s$ background components, $N_i$ is the yield, $P_i$ is the PDF for each component, and the index $j$ indicates the event candidate. The total signal PDF is described as a product of the PDFs for $M_{\rm bc}$ and $\Delta E$. We use the decays $B^+ \to \phi K^+$ and $\bar{B^0} \to \bar{D}^0 \pi^0$ as control samples to correct for differences between data and MC simulations for the fitted means and widths of $M_{\rm{bc}}$ and $\Delta E$. The PDF for $\Delta E$ is a sum of two Gaussians for $B^{+} \to \phi \pi^{+}$ with a common mean, two widths and fraction fixed to the values obtained from a fit to $B^+ \to \phi K^+$ data, and a Crystal Ball function~\cite{CB} with the mean and width fixed to the values derived from $\bar{B^0} \to \bar{D}^0 \pi^0$ data for $B^0 \to \phi \pi^0$. The PDF for $M_{\rm {bc}}$ is a Gaussian function with mean and width fixed to the values obtained from the respective control samples for both modes. To obtain the two-dimensional PDF for the continuum background, we multiply the PDF of $M_{\rm {bc}}$, for which we use an ARGUS~\cite{ARGUS} function, with the PDF of $\Delta E$, which is modeled using a first-order Chebyshev polynomial for $B^{+} \to \phi \pi^{+}$ and $B^{0} \to \phi \pi^{0}$. Both the ARGUS shape parameter and the $\Delta E$ slope are allowed to float. The PDF of the $b \to c$ background is modeled with two-dimensional histograms (2D HistoPDF) with each fixed yield derived from MC simulations. The $b \to u,d,s$ transition backgrounds are modeled with two-dimensional histograms with fixed yields derived from MC simulations except for $B^{+} \to \phi K^{+}$. The PDF for $B^{+} \to \phi K^{+}$ is a double Gaussian function for $\Delta E$ and a Gaussian for $M_{\rm {bc}}$, in which the mean, widths, fraction and yield are fixed to the values derived from a fit to $B^+ \to \phi K^+$ data using the particle indentification requirement $R_{K,\pi } > 0.6$ for kaon candidates. }\par
{Possible backgrounds to $\phi \to K^+ K^-$ decays are predominantly from $B \to K^+ K^- \pi$ with $f_0 (980) \to K^+ K^-$, $a_0 (980) \to K^+ K^-$ or a nonresonant contribution. The two-dimensional fit to $M_{\rm bc}$ and $\Delta E$ alone cannot distinguish the signal from other $B \to K^+K^- \pi$ events. We model $B \to f_0 (980) \pi$, $B \to a_0 (980) \pi$ and nonresonant $B \to K^+ K^- \pi$ with uniform phase space distributions; these backgrounds are treated as additional components in the fits. To evaluate their contributions, we examine events in the $\phi$ mass sidebands, $M_{K^+ K^-} < 1.0$ ${\rm GeV}/c^2$ and 1.039 ${\rm GeV}/c^2$ $<$ $M_{K^+ K^-}$ $<$ 1.1 ${\rm GeV}/c^2$. We apply the same two-dimensional fit to the sideband events assuming that signal-like events are dominated by each of the above three background sources. The possible contribution to the signal is then included as a background PDF corresponding to a signal PDF with fixed mean, width(s) and fraction from each component. As we cannot distinguish these three components, we take the nonresonant mode that gives the largest signal yield as the central value. This background contribution is found to be $4.7 ^{+1.4} _{-1.3}$ events for $B^+ \to \phi \pi^+$ and $ 1.6 ^{+1.0} _{-0.9}$ events for $B^0 \to \phi \pi^0$, derived from the data sideband. The expected yields, 4.7 events for $B^+ \to \phi \pi^+$ and 1.6 events for $B^0 \to \phi \pi^0$, are fixed. We summarize the PDF shape and expected yields (fit outputs) for various components in Table~{\ref{table:pdftable1}}.
\begin{table*}
\caption{\label{table:pdftable1} Summary of the PDF's used in the measurement of $B\to \phi \pi$ decays. Here CB is a Crystal Ball function and 2D HistoPDF is a PDF based on a histogram. Yields in the parentheses are expected values (fit outputs) for the fixed (floated) case.}
\centering{
     \begin{tabular}{l|ll|l|ll|l} \hline
       Mode  &  \multicolumn{3}{|c|} {$B^+ \to \phi \pi^+$} & \multicolumn{3}{|c} {$B^0 \to \phi \pi^0$} \\ \hline

    & $\Delta E$ &  $M_{\rm bc}$      &Method (Yield)          & $\Delta E$  & $M_{\rm bc}$                  & Method (Yield)   \\ \hline \hline
Signal & Sum of two Gaussians & Gaussian & Float($4.5 ^{+5.1} _{-4.3}$) & CB & Gaussian & Float($-2.2 ^{+2.1} _{-1.2}$) \\ 
 $ e^+ e^- \to q \bar{q} $ process  & $1^{st}$ order poly. &ARGUS  &Float($330.0^{+19.1}_{-18.4}$) &$1^{st}$ order poly. & ARGUS &Float($265.6^{+16.9}_{-16.2}$) \\
 $b \to c $  & \multicolumn{2}{c|} {2D HistoPDF}       & Fixed(7.1)  &  \multicolumn{2}{c|}       {2D HistoPDF}        & Fixed(4.8) \\
$ b \to u,d,s $            & \multicolumn{2}{c|} {2D HistoPDF}        & Fixed(4.1)  &\multicolumn{2}{c|} {2D HistoPDF}         & Fixed(13.5) \\
 $B^+ \to \phi K^+$      &Sum of two Gaussians        &Gaussian       & Fixed(33.8)  & -       &-        & - \\
Nonresonant $B \to K^+ K^- \pi$ & Sum of two Gaussians & Gaussian & Fixed(4.7) & CB & Gaussian & Fixed(1.6) \\ \hline \hline
     \end{tabular}
}
\end{table*}
 }\par
{Figure~{\ref{datafit}} shows the $\Delta E$ and $M_{\rm {bc}}$ projections of the fit for the selected $B$ candidates. There are a total of 373 $B^+ \to \phi \pi^+$ and 272 $B^0 \to \phi \pi^0$ candidates in the data sample. We determine the signal yields to be $N_s$($B^{+} \to \phi \pi^{+}$) = $4.5 ^{+5.1} _{-4.3}$ and $N_s$($B^0 \to \phi \pi^0$) = $-2.2 ^{+2.1} _{-1.2}$, where the quoted error is statistical only. We observe no significant signal for $B^{+} \to \phi \pi^{+}$ or $B^0 \to \phi \pi^0$ decays.  
\begin{figure*}
\includegraphics[width=0.83\textwidth]{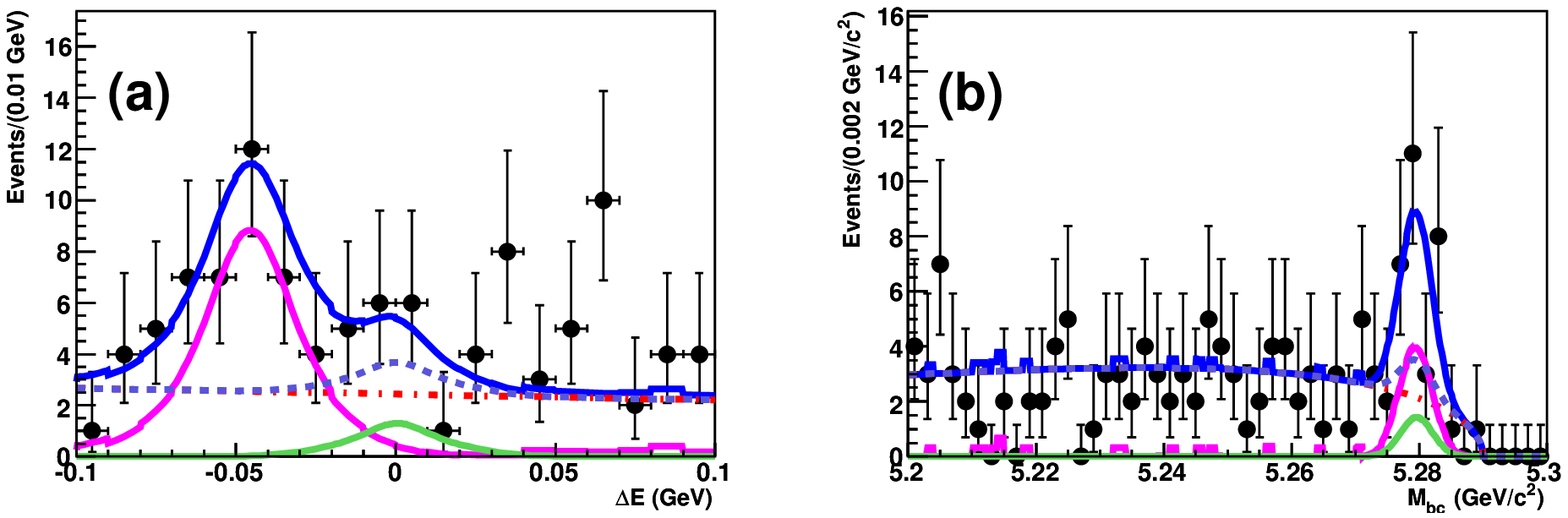}
\includegraphics[width=0.83\textwidth]{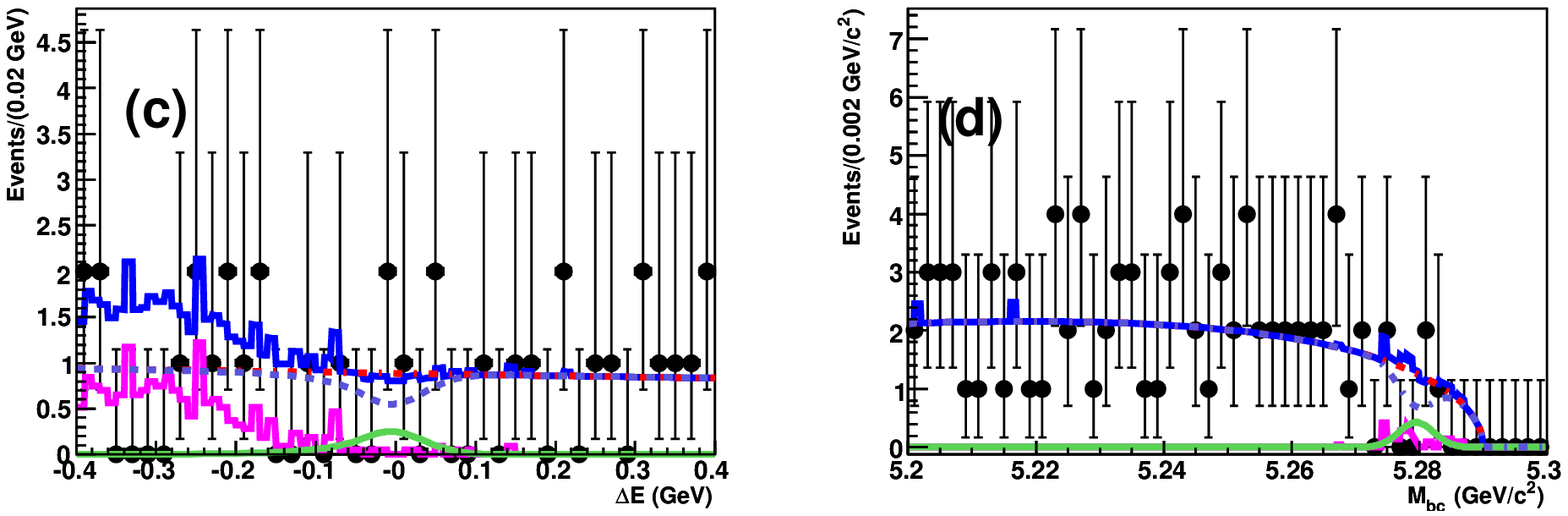}

\caption{Projection of the data (points with error bars) in the fit region. The fit projections onto $\Delta E$ (left) and $M_{\rm {bc}}$ (right) for reconstructed $B^{+} \to \phi \pi^{+}$ (top) and $B^{0} \to \phi \pi^{0}$ (bottom); the sum of signal and $q\bar{q}$ (blue-dotted), $q\bar{q}$ (red-dashed), nonresonant $B \to K^+ K^- \pi$ background (green-solid), other $B$ background (magenta-solid) and the total (blue-solid).
{\bf} }
\label{datafit}
\end{figure*}
The branching fraction $\mathcal B$ is calculated from the observed yield as
\begin{equation}
\mathcal{B}(B \to \phi \pi ) = \frac{ N_{B \to \phi \pi} }{ \epsilon_{\rm data} \times N_{B \overline{B}} }, \\
\end{equation}
where $ N_{B \to \phi \pi}$ is the signal yield, $N_{B \overline{B}}$ is the number of $B \overline{B}$ pairs (where the production rates of $B^+ B^-$ and $B^0 \overline{B^0}$ pairs are assumed to be equal) and $\epsilon_{\rm data}$ is the signal reconstruction efficiency.
The reconstruction efficiency is defined as
\begin{equation}
\quad \epsilon_{\rm data} = \epsilon_{\rm MC} \times \frac{{\epsilon_{\rm data}}^{R_s}}{ {\epsilon_{\rm MC}}^{R_s}} \times \frac{{\epsilon_{\rm data}}^{PID}} {{\epsilon_{\rm MC}}^{PID}} ,\quad
\end{equation}
where $\epsilon_{\rm MC}$ is the reconstruction efficiency from MC simulations and the branching fractions, $\mathcal{B}({\phi \to K^+ K^-}) = 48.9\% $ and $\mathcal{B}({\pi^0 \to \gamma \gamma}) = 98.8\%$, are applied to MC simulations. ${\epsilon_{\rm data}}^{R_s}$ (${\epsilon_{\rm MC}}^{R_s}$) is the efficiency of the ${R_s}$ requirement from data (MC simulations), ${\epsilon_{\rm data}}^{PID}$ (${\epsilon_{\rm MC}}^{PID}$) is the efficiency of the PID requirement from data (MC simulations). }\par
\begin{table}[htb]
\caption{Summary of systematic uncertainties (events) in the signal yield ($Y$) extraction. }
\label{sys_fit}
\centering{
     \begin{tabular}{c|cc} \hline \hline
Sources            & $B^{+} \to \phi \pi^{+}$      & $B^0 \to \phi \pi^0$ \\\hline\hline
Signal PDF & $ ^{+0.5} _{-0.6}$   & $ ^{+0.6} _{-0.4}$   \\
$b \to u,d,s$ &$^{+0.1}_{-0.1} $  & $^{+0.0}_{-0.1} $     \\
$B^+ \to \phi K^+$     &${^{+1.8} _{-1.6}} $  & $-$   \\
nonresonant $B \to K^+ K^- \pi$     &${\pm 2.4} $  & ${\pm 0.8} $ \\ 
Fit bias     &$ ^{+0.9} _{-0.0}$   & $ ^{+0.3} _{-0.0}$  \\
Peaking background modeling     & $ ^{+0.0} _{-6.3}$   & $ ^{+0.0} _{-2.2}$ \\\hline\hline
Total              &  $^{+3.1} _{-6.9} $   &  $^{+1.3} _{-2.4}$  \\\hline\hline
     \end{tabular}
}
\end{table} 
{We consider the systematic uncertainties in the efficiency, $N_{B\bar{B}}$ and the yield extraction. The main sources of efficiency uncertainties are MC statistics $0.6\%$ $(0.8\%)$, PID $2.0\%$ $(1.3\%)$ and tracking $3.1\%$ $(2.0\%)$ for $B^+ \to \phi \pi^+$ $(B^0 \to \phi \pi^0)$. The uncertainty on the $\pi^0$ efficiency is measured by comparing the yields between $\eta \to \gamma \gamma $ and $\eta \to \pi^0 \pi^0 \pi^0$ and is found to be 3.0\%. To evaluate the uncertainty from the efficiencies due to the $R_S$ requirements, we use the control samples $B^+ \to \bar{D^0} (\bar {D^0} \to K^+ \pi^-) \pi^+$ for $B^+ \to \phi \pi^+$ and $B^0 \to D^+ (D^+ \to K^0 _S \pi^+) \pi^-$ for $B^0 \to \phi \pi^0$. The $R_S$ uncertainties are 2.4\% (4.1\%) for $B^+ \to \phi \pi^+$ $(B^0 \to \phi \pi^0)$. The uncertainty from $N_{B\bar{B}}$ is 1.4\%. The sources and sizes of systematic uncertainies in the signal yield extraction are listed in Table~{\ref{sys_fit}}. The systematic error from signal yield extraction is estimated by varying all fixed parameters by $\pm 1 \sigma$. To obtain the errors due to the fixed yields of $b \to u,d,s$ backgrounds, $b \to c$ backgrounds and nonresonant $B \to K^+ K^- \pi$, we vary these fixed yields by $\pm 50\%$. The uncertainty from the $b \to c$ backgrounds is negligible. The largest difference in yield between nonresonant $B \to K^+ K^- \pi$ and the other modes is included in the systematic error. The uncertainty from this difference, which is the largest contributor to the total systematic error, is $-6.3$ $(-2.2)$ events for $B^+ \to \phi \pi^+$ $(B^0 \to \phi \pi^0 )$.}\par 
{The upper limit (${\cal B}_{UL}$) is determined as 
\begin{equation}
{{ \int _0 ^{\mathcal{B}_{UL}} \mathcal{L(\mathcal{B})}d\mathcal{B}} \over { \int _0 ^{\infty} \mathcal{L(\mathcal{B})}d\mathcal{B}}} =0.90,
\end{equation}
where $\mathcal{L(\mathcal{B})}$ is the likelihood value and $\mathcal{B}$ is the branching fraction. The branching fraction is determined as the number of the signal events divided by the number of $B \bar{B}$ pairs and the reconstruction efficiency. We include systematic errors by convolving the likelihood function with a Gaussian whose width is equal to the total systematic error. The upper limits on the branching fractions are found to be ${\cal{B}}(B^{+} \to \phi \pi^{+})$ $<$ $3.3 \times 10^{-7}$ and ${\cal{B}}(B^0 \to \phi \pi^0)$ $<$ $1.5 \times 10^{-7}$ at the 90\% CL. The results, together with the central values for the branching fractions, are listed in Table~\ref{br}. }\par
\begin{table}[htb]
\caption{Signal yields, measured branching fractions including statistical and systematic errors, and the upper limits including systematic uncertainties at the 90\% CL.}
\label{br}
\centering{
     \begin{tabular}{ccc} \hline \hline
            &  $B^{+} \to \phi \pi^{+}$\    & $B^0 \to \phi \pi^0$ \\\hline
Yield       &$4.5 ^{+5.1} _{-4.3} {^{+3.1} _{-6.9}}$  &$-2.2 ^{+2.1} _{-1.2} {^{+1.3} _{-2.4}}$   \\
$\epsilon_{\rm data}$ &8.4\%  &4.9\%  \\
$\mathcal{B}(10^{-7})$  & $0.8 ^{+0.9} _{-0.8} {}^{+0.6} _{-1.3}$  &$ -0.7 ^{+0.6} _{-0.4} {}^{+0.4} _{-0.8}$ \\
${\cal B}_{UL}(10^{-7})$ &$ 3.3$ &$ 1.5$        \\\hline\hline
     \end{tabular}
}
\end{table}  
{In summary, using $657 \times 10^6$ $B \overline{B}$ pairs collected at the $\Upsilon(4S)$ with the Belle experiment, we find no significant signals for $B^{+} \to \phi \pi^{+}$ and $B^0 \to \phi \pi^0$. We set upper limits of $\mathcal{B}(B^{+} \to \phi \pi^{+}) < 3.3 \times 10^{-7}$ and $\mathcal{B}( B^0 \to \phi \pi^0 ) <1.5 \times 10^{-7}$ at the 90\% CL.}

We thank the KEKB group for excellent operation of the
accelerator; the KEK cryogenics group for efficient solenoid
operations; and the KEK computer group, the NII, and 
PNNL/EMSL for valuable computing and SINET4 network support.  
We acknowledge support from MEXT, JSPS and Nagoya's TLPRC (Japan);
ARC and DIISR (Australia); NSFC (China); MSMT (Czechia);
DST (India); INFN (Italy); MEST, NRF, GSDC of KISTI, and WCU (Korea); 
MNiSW (Poland); MES and RFAAE (Russia); ARRS (Slovenia); 
SNSF (Switzerland); NSC and MOE (Taiwan); and DOE and NSF (USA).


\begin{thebibliography}{99}
\bibitem{conjugate}
Inclusion of the charge-conjugate state is implied throughout this paper.
\end{thebibliography}
\end{document}